\documentclass{aip-cp}

\usepackage[numbers]{natbib}
\usepackage{rotating}
\usepackage{graphicx}

\begin{document}

\title{Study and characterization of SrTiO$_{3}$ surface}

\author[aff1,aff2]{Fatima Alarab}
\author[aff2]{J\'an Min\'ar}
\author[aff2]{Pavol \v{S}utta}
\author[aff2]{Lucie Pru\v{s}\'akov\'a}
\author[aff2]{Rostislav Medl\'in}
\author[aff1]{Olivier Heckmann}
\author[aff1]{Christine Richter}
\author[aff1]{Karol Hricovini}

\affil[aff1]{LPMS,Cergy-Pontoise University, Neuville, France}
\affil[aff2]{New Technologies Research Centre, University of West Bohemia, Plzen, Czech Republic}

\maketitle
\begin{abstract}
The two-dimensional electron gas (2DEG) at oxides interfaces and surfaces has attracted large attention in physics and research due to its unique electronic properties and possible application in optoelectronics and nanoelectronics. The origin of 2DEGes at oxide interfaces has been attributed to the well known "polar catastrophe" mechanism. On the other
hand, recently a 2DEG was also found on a clean SrTiO$_{3}$(001) surface where it is formed due to oxygen vacancies. However, these 2DEG systems have been until now found mostly on atomically perfect crystalline samples usually grown by pulsed laser deposition or molecular beam epitaxy i.e. samples which are difficult to be prepared and require specific experimental conditions. Here, we report on the fabrication of SrTiO$_{3}$ thin films deposited by magnetron sputtering which is suitable for mass-production of samples adapted for nanoelectronic applications. The characterisation of their structural and electronic properties was done and compared to those of SrTiO$_{3}$ single crystals. XRD patterns and SEM micrography show that the deposited films are amorphous and their structure changes to polycrystalline by heating them at 900$^{\circ}$C. Photoemission spectroscopy (XPS and UPS) was used to study the electronic properties of the films and the crystal. In both, we observe the 2DEG system at Fermi level and the formation of Ti$^{3+}$ states after heating the surface at 900$^{\circ}$C. 
\end{abstract}

\section{INTRODUCTION}
The increase of the efficiency of solar cells was always one of the top goals in photovoltaic applications and has enhanced the activity of studies on perovskite thin films. Transition metal oxides with perovskite structure are of interest because of their powerful properties such as high dielectric constant, magneto-electric coupling, chemical stability and photo-activity. Indeed, SrTiO$_{3}$ is one of the titanates that is considered a good candidate for dielectric and photoelectric applications. SrTiO$_{3}$ thin films can be found in a wide range of applications, in particular in solar cells. 
\\According to previous studies, the intrinsic and fundamental properties of transition metal oxides are strongly related to the experimental conditions of growth and surface treatment. This is due to the role that defects play in these materials since they cause disorder and affect the functionality of the hosting system. Moreover, despite the insulating nature of SrTiO$_{3}$, this latter has shown the possibility to create a 2-dimensional electron gas (2DEG) on its surface and at the interfaces with other oxides. 
\\One example is the famous crystalline LaAlO$_{3}$/SrTiO$_{3}$ heterostructure discovered for the first time by A. Ohtomo and H.Y. Hwang \cite{bib1}. They report the formation of a 2DEG at this interface and the resulting properties such as superconductivity and magnetism. Others also revealed that a large number of electrons were confined at the heterojunction of LaAlO$_{3}$ and SrTiO$_{3}$ layers due to the well-known polar catastrophe mechanism \cite{Nakagawa} in the origin of 2DEG. Thus, a series of experimental studies were performed and theoretical investigations as well for a better understanding of the 2DEG formation at LaAlO$_{3}$/SrTiO$_{3}$ interfaces \cite{Hwang,Huijben,Popovic,Kalabukhov,Willmott,Pentcheva,Ariando}. 
\\Another type of 2DEG systems is found at amorphous/crystalline interfaces \cite{bib2}. The two types have similar physical properties in terms of superconductivity, potential-well depth, etc. \cite{bib3,bib4,bib5}, but also many differences as well, e.g. the 2DEG has less lattice strain at the interface, higher interface carrier density and fewer requirements for growth for this second type.
\\Oxygen vacancies are considered as another possible origin of electron sources at the interfaces \cite{Eckstein,Zhong} and the main reason in the formation of a 2DEG at the surface of SrTiO$_{3}$ \cite{bib6,bib7}. Several studies on the origins of oxygen vacancies and their role in the formation of 2DEG have been reported \cite{bib3,bib8,bib9}. However, according to literature, almost all the studies that were carried out about the formation of a 2DEG at the SrTiO$_{3}$ surface have been done on samples with very good crystalline quality grown by Pulsed Laser Deposition (PLD) or Molecular Beam Epitaxy (MBE). Yet, all these techniques require specific growth conditions such as high temperature ( 600$^{\circ}$C), crystalline substrate etc, they are time-consuming and expensive as well which is not suitable for applications from the economical point of view.     
Starting from these reasons, many studies have carried out on the properties of SrTiO$_{3}$ thin films by using other growth techniques. Dubourdieu has reported the dielectric properties of SrTiO$_{3}$ thin films deposited by MOCVD with respect to thickness and composition \cite{bib10}, S H Nam et al discussed the electrical properties of SrTiO$_{3}$ thin films deposited on a Si(100) substrate by RF magnetron sputtering \cite{bib11}. 
\\In this work, we aim to study the quality of SrTiO$_{3}$ films deposited on crystalline Si substrate terminated with amorphous SiO$_{2}$ layer by RF magnetron sputtering, their structure and composition by comparing them with the surface state of single crystalline SrTiO$_{3}$(100).

\section{Experimental methods}
The films were prepared by RF (13.56 MHz) magnetron sputtering using a BOC Edwards TF 600 coating system. Before the films growth, the deposition chamber was evacuated to a base pressure of 2.10$^{-4}$ Pa. The substrates were cleaned by ion etching in argon plasma by applying a RF power of 200 W for 15 minutes at 0.2 Pa. A pure target of SrTiO$_{3}$ was placed on a magnetron connected to a RF power supply.The deposition was kept under constant discharge RF power (400 W). The films were deposited on amorphous Si substrates at 400$^{\circ}$C in argon atmosphere at constant pressure 0.6 Pa. The thickness of the films is in the range of 70 to 100 nm.
\\The structure of the films was studied by X-Ray diffraction (XRD) using an automatic powder diffractometer X’Pert  Pro with CuK$\alpha$ radiation. XRD patterns were collected using the asymmetric $\omega$-2$\theta$ geometry with $\omega$=0.5$^{\circ}$ from 15 to 70 degrees in 2$\theta$ scales \cite{bib13}. A line profile analysis of the strongest lines was performed in order to calculate the crystallite sizes of the films by procedures based on a Voigt function \cite{bib14,bib15}.
\\The characterisation of the surface morphology of the films was carried out using the scanning electron microscopy JOEL JSM 7600F operated at 30 kV (field emission (Schottky) with a resolution of 1 nm at 15 kV) with an Energy Dispersive X-ray (EDX) SDD detector Oxford Instruments X-Max attached. 
\\We characterised electronic properties by photoemission. Core levels were measured by X-Ray photoemission spectroscopy (XPS) using Mg K$\alpha_{1,2}$ radiation ($\hbar\nu$=1256 eV) and the valence band was studied by Ultraviolet Photoemission Spectroscopy (UPS) with HeII radiation ($\hbar\nu$=40.8 eV).
In the photoemission experiment one needs to distinguishes between two different approaches. Angle-resolved measurements are possible for single crystal samples because they allow to access the $k$-vector of the initial state. Angle integrated measurements are used for polycrystalline samples \cite{bib12}, as in this case $k$-vector information is blurred by integration over all electronic states in the Brillouin zone.

\begin{figure}[h!]
	\centerline{\includegraphics[width=360pt]{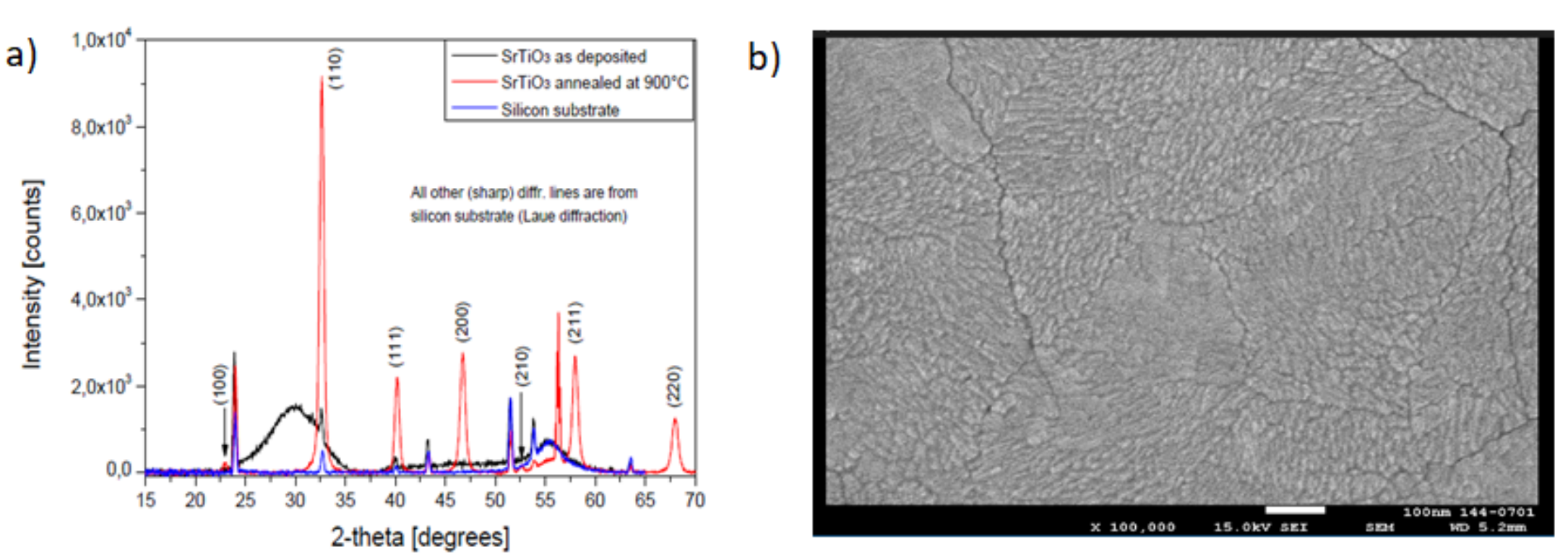}}
	\caption{a) XRD patterns of a STO thin film after deposition (black) and after heating at 900$^{\circ}$C (red). The blue represents the diffraction patterns of the naked silicon substrate. b) SEM micrography of the STO thin film after heating at 900$^{\circ}$C }
\end{figure}

\section{Results and discussion}
\subsection{1. Structural and surface properties}
The XRD patterns of the as-deposited and annealed SrTiO$_{3}$ films are represented in Figure 1(a). The broad lines of as-deposited SrTiO$_{3}$ show that the films are amorphous. The amorphous nature of the as-deposited films is due to the growth method and the experimental conditions during the deposition at room temperature. After heating at high temperature (900$^{\circ}$C), these lines change to sharp lines corresponding to a polycrystalline structure of SrTiO$_{3}$ with a preferred orientation in the (110) direction. The average dimension of the diffracting domains (crystallites) is about 50 to 60 nm. Figure 1(b) illustrates the SEM micrography of the annealed SrTiO$_{3}$ films. It can be seen that the surface of these films which have polycrystalline structure is rough and has grain boundaries. The SEM image also confirms the size of the crystallites obtained from the XRD data.

\subsection{2. Electronic properties}
 
Due to a short mean free path of photoelectrons (typically 1 nm at the electron kinetic energy of about 40 eV), the photoemission measures only top most layers of a sample. So, a cleaning procedure is to be applied to remove a contamination film sitting on the surface. We applied the same approach of surface preparation for both, the polycristalline films and the crystal: 
\\i. After the transfer into the vacuum the samples were cleaned by heating them at 400$^{\circ}$C (which is the same temperature that was used during the RF film deposition) at a base pressure of 2.10$^{-9}$ mbar for 3 hours, 
\\ii. Samples were then heated at 900$^{\circ}$C in a vacuum of P=5.10$^{-9}$ mbar for 5 hours.
\begin{figure}[h!]
	\centerline{\includegraphics[width=350pt]{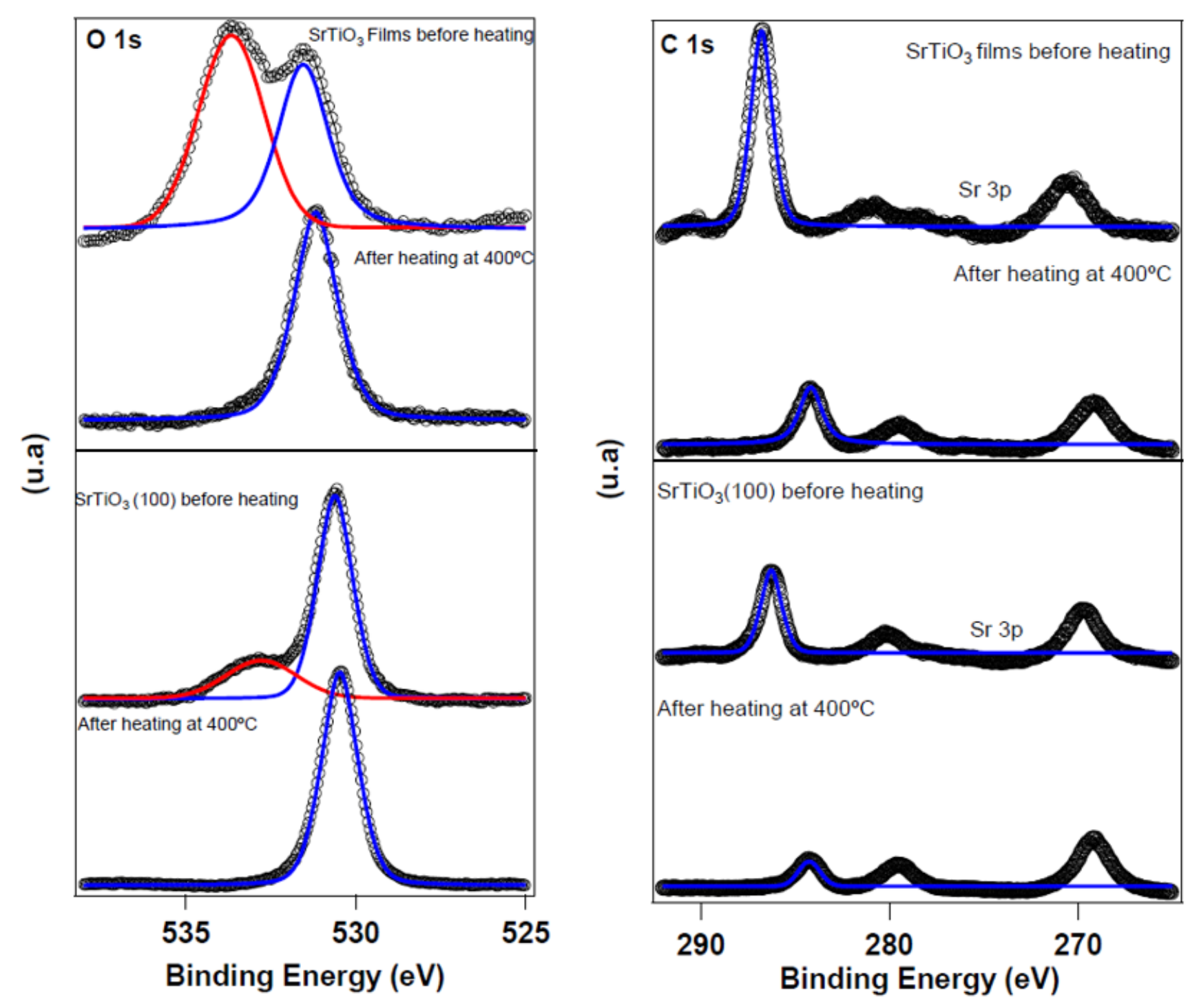}}
		\caption{a) O 1s and b) C 1s core levels of SrTiO$_{3}$ films and crystal before and after they were heated at 400$^{\circ}$C in the vacuum, respectively.}
\end{figure}
\\O 1s and C 1s core levels were first measured, after introducing the film and the crystal sample into the vacuum and again after heating them at 400$^{\circ}$C. The resulting spectra are represented in Fig.2(a) for O 1s and Fig.2(b) for C 1s core levels, respectively. Before heating at 400$^{\circ}$C, O 1s spectra (Fig.2(a)) show two peaks around 531 eV and 533 eV while only the main peak at 531 eV is observed after heating at 400$^{\circ}$C in the spectra of both samples. In Fig.2(b), C 1s spectra show an intense peak at 286.5 eV for film and crystal samples after introduction into the vacuum. The intensity of the peak is considerably reduced and the peak is shifted to 284.8 eV after the cleaning procedure at 400$^{\circ}$C. When samples are exposed to the air, the main contamination is formed by hydrocarbons and by C-O, C=O radicals. The heating procedure removes a majority of the contaminants. Clearly, after heating, only one component of O 1s is detected corresponding to oxygen in SrTiO$_{3}$. This is corroborated by the reduced C 1s signal. However, a presence of residual and shifted carbon signal is still detected on the cleaned surface. It testifies that the carbon atoms are presumably forming sp$^{3}$ and/or sp$^{2}$ bonds, a signature of graphitic-like compounds. In the case of SrTiO$_{3}$ films, carbon can be present in the bulk of the samples, as the preparation is performed in rather low vacuum (2.10$^{-4}$ Pa) conditions, explaining why its contribution is smaller in the spectrum than for the other sample.

\begin{figure}[h!]
	\centerline{\includegraphics[width=350pt]{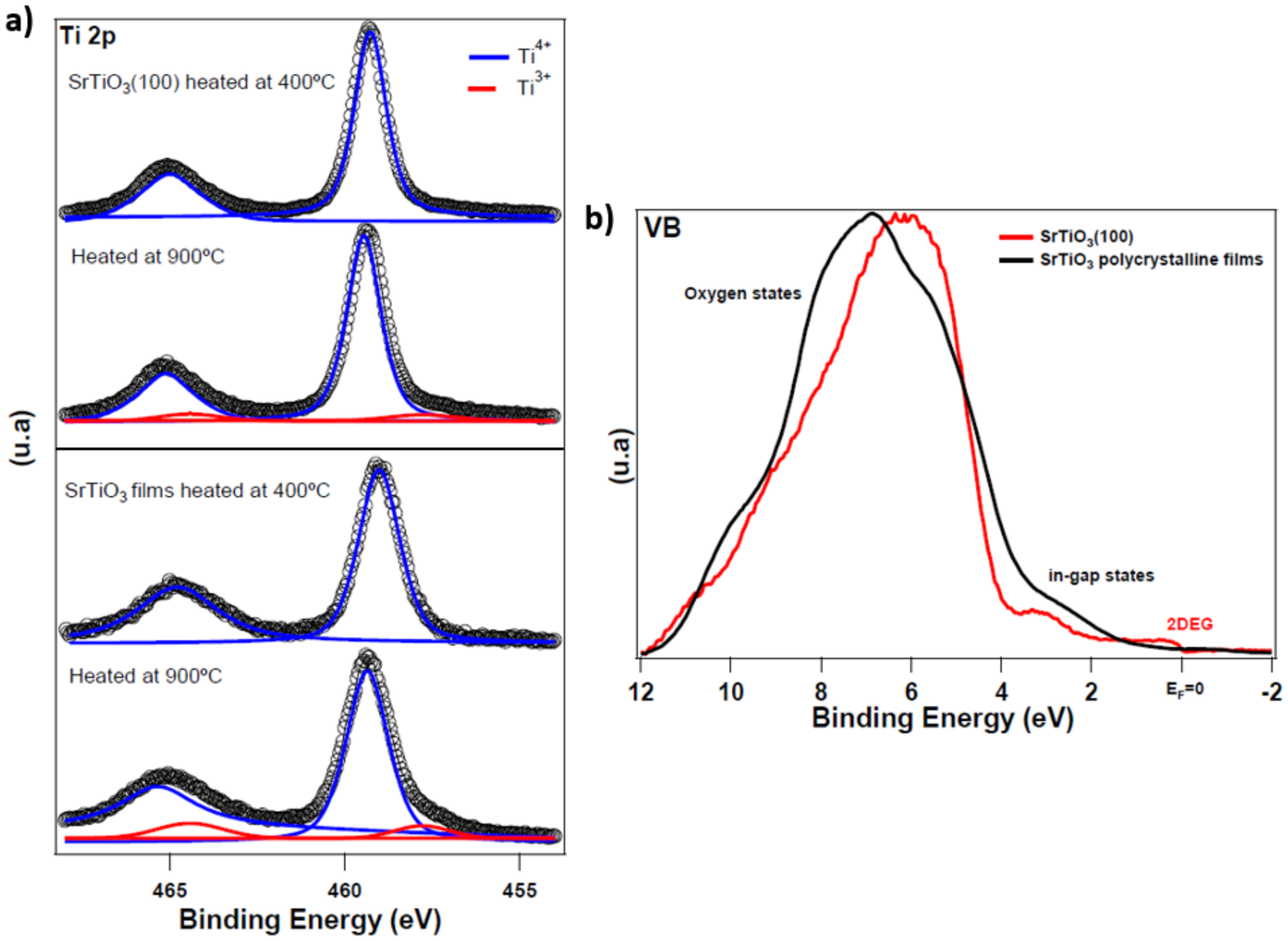}}
	\caption{a)Ti 2p core levels of SrTiO$_{3}$ films and crystal before and after they were heated at 900$^{\circ}$C in the vacuum. The corresponding fits of the spectra describe different valence states: Ti$^{4+}$ in blue for and for Ti$^{3+}$ in red. b) Valence band of SrTiO$_{3}$ polycrystalline films (blue) and crystal (red) after they were heated at 900$^{\circ}$C in the vacuum, representing the 2DEG at Fermi level, in-gap states and oxygen bands.}  
\end{figure}                   
Figure 3(a) shows Ti 2p core levels measured by XPS for SrTiO$_{3}$ films and single crystal respectively, heated  at 400$^{\circ}$C and then annealed at 900$^{\circ}$C for 5 hours. In the following, only the Ti 2p$_{3/2}$ peak will be discussed since it is more intense than the Ti 2p$_{1/2}$ peak. As can be seen in the figure, for both samples, film and crystal, after heating at 400$^{\circ}$C, Ti 2p$_{3/2}$ shows a single peak at 459.5 eV which corresponds to the Ti$^{4+}$ valence state existing in SrTiO$_{3}$. After baking at high temperature (900$^{\circ}$C), a low binding energy shoulder appears in the Ti 2p$_{3/2}$ spectra at 457.7 eV in both samples, film and single crystal as well. Note that the film has a polycrystalline structure now, as evidenced and discussed in Figure 1. The energy difference between these two states is about 1.8 eV which is equal to the chemical shift between Ti$^{4+}$ and Ti$^{3+}$. The same result was reported by many previous studies on SrTiO$_{3}$ single crystals. According to references \cite{bib8,bib16} the formation of Ti$^{3+}$ represents a clear signature of a reduced surface, consistent with the formation of a metallic-like Fermi edge. 
To test this hypothesis on polycrystalline films, we measured the valence band at the surface of films and single crystal SrTiO$_{3}$(100) after they were heated at high temperature of 900$^{\circ}$C in vacuum. 
\\Figure 3(b) shows valence band spectra measured with a photon energy of 40.8 eV. The main structure between 4 and 8 eV represents the O 2p states. The band gap of SrTiO$_{3}$ has a value of about $\approx$3.2 eV \cite{bib6,bib8}. The gap cannot be identified in our spectrum because of in-gap states that exist on both film and crystal surfaces between 1.5 and 3.5 eV. Other important features in these spectra are the 2DEG peak at Fermi level clearly observed on the surface of the single crystal, on the polycrystalline sample it cannot be completely excluded but it definitely is much smaller. According to previous experimental findings, in-gap states are related to oxygen vacancies and are detected between 1 and 1.5 eV for samples measured at low and room temperature \cite{bib6,bib8,bib9}. Theoretical models predict these states at binding energies between 2 and 4 eV \cite{bib17}. The distribution of these states and their origins depend strongly on the growth method, the structure and the surface state of the sample, which can lead to different binding energies and/or intensities. 
\\As deduced from XRD patterns and SEM images, the films evolve from amorphous to polycrystalline structure when they are heated at 900$^{\circ}$C with a preferred orientation along (110) direction. Since the size of crystallites is relatively small (60 nm) compared to the diameter of our photon beam (0.5 mm), during the photoemission measurement we integrate over a large number of crystallites, mostly in the (110) direction. Y. Aiura et al. in reference \cite{Y} reported the formation of 2DEG at Fermi level and some states in the gap for SrTiO$_{3}$(110) heated at 800$^{\circ}$C. The intensity of this 2DEG system became very weak after they heated to 1000$^{\circ}$C and only one state at 1.1 eV was identified in the gap. They related this observation directly to the surface structural change. In our experiments, the polycrystalline films were heated to 900$^{\circ}$C and all the crystallites do not have the same orientation. This may explain a very weak intensity at Fermi level, corresponding to the 2DEG peak, in our spectrum.  
\section{Conclusion}
Our observations show that the structure of SrTiO$_{3}$ films deposited by magnetron sputtering is amorphous and changes to polycrystalline when the films are heated at high temperature (900$^{\circ}$C). We have used photoemission to observe the formation of a 2DEG at the surface of SrTiO$_{3}$ polycrystalline films and of a single crystal at room temperature after they were annealed at 900$^{\circ}$C. Our experimental findings indicate that the structure and electronic properties of SrTiO$_{3}$ surface are highly dependent on experimental preparation conditions. Namely, they determine formation of oxygen vacancies which are considered to be an important factor in the formation of 2DEG at the SrTiO$_{3}$ surface. We found in gap states in polycrystalline films, probably referred to oxygen vacancies, similar to those of a single crystal.     
\section{ACKNOWLEDGMENTS}
This work was supported
by the CEDAMNF project (CZ.02.1.01/0.0/0.0/15$\_$003/0000358), co-funded
by the ERDF as part of the OP RDE program of the Ministry of Education,
Youth and Sports (Czech Republic).

\end{document}